# Borel Resummation Method with Conformal Mapping and the Ground State Energy of the Quartic Anharmonic Oscillator

*Wajdi Abdul Aziz Gaddah* and *Ibrahim Saleh Jwan*

*Department of Physics, Al Fatah University, P.O.Box* 84441, *Tripoli, Libya*

wajdigaddah@yahoo.co.uk

*Abstract*: In this paper, we consider the resummation of the divergent Rayleigh-Shrödinger perturbation expansion for the ground state energy of the quartic anharmonic oscillator in one dimension. We apply the Borel-Padé resummation method combined with a conformal mapping of the Borel plane to improve the accuracy and to enlarge the convergence domain of the perturbative expansion. This technique was recently used in perturbative QCD to accelerate the convergence of Borel-summed Green's functions. In this framework, we calculated the ground state energy of the quartic anharmonic oscillator for various coupling constants and compared our results with the ones we obtained from the diagonal Padé approximation and the standard Borel resummation technique. The results are also tested on a number of exact numerical solutions available for weak and strong coupling constants. As a part of our calculations, we computed the coefficients of the first 50 correction terms in the Rayleigh-Shrödinger perturbation expansion using the method of Dalgarno and Stewart. The conformal mapping of the Borel plane is shown to enhance the power of Borel's method of summability, especially in the strong coupling domain where perturbation theory is not applicable.

## I. INTRODUCTION

In quantum mechanics, exact analytical solutions for the Schrödinger wave equation can be obtained only for a few idealized systems such as the hydrogen atom and the simple harmonic oscillator. For physical systems described by complicated Hamiltonians one therefore has to resort to numerical techniques or approximation methods. Perturbation theory in its region of applicability provides a useful tool for estimating eigenvalues and eigenfunctions for physical problems whose exact solutions can not be obtained analytically. Rayleigh-Shrödinger (RS) perturbation theory [1] is one of the few principal methods for approximating the energy eigenvalues and eigenfunctions of a Hamiltonian which can be partitioned into a completely solvable part and a small perturbation. The lack of an exact analytical solution to the eigenvalue problem of the quartic anharmonic oscillator in quantum mechanics makes this problem one of the most popular theoretical laboratories for examining the validity of various approximation techniques. Bender and Wu [2], [3] showed that the RS perturbative expansions for the energy eigenvalues of the quartic anharmonic oscillator diverge for all values of the coupling constant $\lambda$, i.e. they have a zero radius of convergence. Therefore, resummation procedures such as Padé approximation [4], [5] are necessary to extract physical results from perturbatively divergent power series expansions. One method of resummation that has been of interest in quantum field theory is the Borel's method of summability [5], [6] refined by conformal mapping of the Borel plane [7]-[10]. Mansfield in [11] proposed a modified type of Borel resummation in the form of the famous Cauchy integral formula. His method depends on a free parameter that he introduced to cancel some singularity contributions to his contour integral. Although Mansfield managed to calculate the ground state energy of the quartic anharmonic oscillator his method still suffers from the lack of a precisely standard criterion to fix this parameter. The aim of this paper is to explore and enhance Borel's summability with the conformal mapping technique without introducing any free parameters and then use this method in quantum mechanics to resum the divergent RS perturbative expansion for the ground state energy of the quartic anharmonic oscillator in weak, intermediate and moderately strong coupling domains.

## II. THE GROUND-STATE ENERGY OF THE QUARTIC ANHARMONIC OSCILLATOR IN RS PERTURBATION THEORY

In this section, we illustrate within the framework of RS perturbation theory the method of Dalgarno and Stewart [12] in calculating the ground-state energy $E_0$ of the quartic anharmonic oscillator. In dimensionless form, the time-independent Schrodinger equation for the one-dimensional quartic anharmonic oscillator reads as:

$$(-\frac{d^2}{dx^2} + x^2 + \lambda x^4)\psi_n(x) = E_n \psi_n(x), \quad (1)$$

where $\lambda$ is a dimensionless positive coupling constant governing the strength of perturbation and $\psi_n(x)$ is an energy eigenfunction obeying the boundary conditions $\psi_n(\pm\infty) = 0$. In RS perturbation theory, we represent $E_0$ and $\psi_0(x)$ as power series in $\lambda$:

$$E_0(\lambda) = \sum_{n=0}^{\infty} E_0^{(n)} \lambda^n \quad \text{and} \quad \psi_0(x) = \sum_{n=0}^{\infty} \lambda^n u_0^{(n)}(x). \quad (2)$$

Since (1) is linear we can adopt the normalization condition $\psi_0(0) = 1$, which translates into $u_0^{(0)}(0) = 1$ and $u_0^{(n)}(0) = 0$ for all $n > 0$. The unperturbed ground-state energy eigenvalue $E_0^{(0)}$ and eigenfunction $u_0^{(0)}(x)$ in (2) are given by:

$$E_0^{(0)} = 1 \quad \text{and} \quad u_0^{(0)}(x) = e^{-x^2/2}. \tag{3}$$

The perturbation corrections $u_0^{(n)}(x)$ and the expansion coefficients $E_0^{(n)}$ in (2) can be determined in a simple and more practical method developed by Dalgarno and Stewart [12]. In this approach, we express $u_0^{(n)}(x)$ as the product of $u_0^{(0)}(x)$ in (3) and a polynomial $F_n(x)$:

$$u_0^{(n)}(x) = e^{-x^2/2} F_n(x), \tag{4}$$

where $F_0(x) = 1$ and $F_n(0) = 0$ for all $n > 0$. Substituting (2) into (1), we obtain with the help of (4) the following recursion relation for $F_n(x)$:

$$2x F_n'(x) - F_n''(x) + x^4 F_{n-1}(x) = \sum_{k=1}^{n} E_0^{(k)} F_{n-k}(x), \tag{5}$$

here $n$ is the order of perturbation. Inspection of (5) reveals that the polynomial solutions $F_n(x)$ are of the form [2]:

$$F_n(x) = \sum_{k=1}^{2n} b_k^{(n)} x^{2k}, \quad n = 1, 2, 3, \ldots. \tag{6}$$

By substituting (6) into (5) we obtain a recursion relation for the coefficients $b_k^{(n)}$ and a simple formula for $E_0^{(n)}$:

$$E_0^{(n)} = -2 b_1^{(n)}, \tag{7}$$

$$(2k+1)(2k+2) b_{k+1}^{(n)} = 4k b_k^{(n)} + b_{k-2}^{(n-1)} + 2 \sum_{r=[k/2]}^{n-1} b_1^{(n-r)} b_k^{(r)}, \tag{8}$$

where $k = 1, 2, 3, \ldots, 2n$ and $[k/2] = k/2$ for $k$ even and $(k+1)/2$ for $k$ odd. In (8) we use the convention that $b_{2n+1}^{(n)} = b_{-1}^{(n-1)} = 0$ and $b_0^{(n-1)} = \delta_{0, n-1}$, where $\delta_{n,m}$ is the Kronecker delta. From (7) and (8) we computed by the exact rational arithmetics of Maple the coefficients $E_0^{(n)}$ of the first 50 terms in the RS expansion (2) and listed the first 30 of them in Table IV for illustrating their divergent behaviour only. Now, with the knowledge of the expansion coefficients $E_0^{(n)}$ we obtain the RS perturbative expansion for $E_0(\lambda)$:

$$E_0(\lambda) = \sum_{n=0}^{\infty} E_0^{(n)} \lambda^n = 1 + \frac{3}{4} \lambda - \frac{21}{16} \lambda^2 + \frac{333}{64} \lambda^3 \\ - \frac{30885}{1024} \lambda^4 + \frac{916731}{4096} \lambda^5 - \frac{65518401}{32768} \lambda^6 + \cdots. \tag{9}$$

This expansion has a zero radius of convergence as can be seen from the large-order asymptotic form of the coefficients $E_0^{(n)}$, which was found by Bender and Wu [2], [13] to be:

$$E_0^{(n)} \simeq -r (-1)^n (3/2)^n (n - 1/2)! \quad \text{as} \quad n \to \infty, \tag{10}$$

where $r > 0$. In what follows, we denote by $E_{RS}^{(N)}(\lambda)$ the $N$th partial sum of the RS series (9). As the order of perturbation $N$ increases considerably the partial sum $E_{RS}^{(N)}(\lambda)$ tends to diverge even for small values of $\lambda$. Table I shows the rate of divergence of $E_{RS}^{(N)}(\lambda)$ with increasing $N$ for relatively small $\lambda$. For this reason, resummation procedures such as Padé approximant and Borel method are necessary for extracting physical results from divergent series in perturbation theory.

### III. Padé Approximation to the RS expansion for $E_0(\lambda)$

In this section, we apply Padé approximant (PA) [4], [5] to resum the divergent asymptotic series (9), representing the ground-state energy in RS perturbation theory. The Padé approximant of order $[m,n]$ for the perturbative expansion (9) is a rational function of the form:

$$E_{PA}^{[m,n]}(\lambda) = \frac{a_0 + a_1 \lambda + a_2 \lambda^2 + \cdots + a_m \lambda^m}{1 + b_1 \lambda + b_2 \lambda^2 + \cdots + b_n \lambda^n}. \tag{11}$$

More accuracy is achieved by choosing $m = n$ or $m = n + 1$, [15], [16]. Padé coefficients $a_i$ and $b_i$ in (11) can be determined uniquely from the condition:

$$E_0(\lambda) - E_{PA}^{[m,n]}(\lambda) = O(\lambda^{m+n+1}), \quad \lambda \to 0, \tag{12}$$

which gives for $m \geq n$ a set of $m + n + 1$ linear equations:

$$\sum_{i=0}^{k} b_i E_0^{(k-i)} = a_k, \quad k = 0, 1, \ldots, m,$$

$$\sum_{i=0}^{n} b_i E_0^{(k-i)} = 0, \quad k = m+1, \ldots, m+n, \tag{13}$$

where $b_0 = 1$ and $b_i = 0$ for $i > n$. By solving (13) we obtain the $(m + n + 1)$ unknowns $\{a_0, a_1, \ldots, a_m, b_1, \ldots b_n\}$ in terms of the energy coefficients $\{E_0^{(0)}, E_0^{(1)}, \ldots, E_0^{(m+n)}\}$. From now on, we shall work only with the diagonal $[N, N]$ Padé approximants because they usually converge faster with increasing $N$, [13], where $2N$ is the order of perturbation. Unlike the RS perturbative expansion, the $[N, N]$ Padé approximant for small and relatively large $\lambda$ converges to the exact result [13] as $N \to \infty$. Tables I and II show the rate of convergence of $E_{PA}^{[N,N]}(\lambda)$ with increasing $N$ for different values of $\lambda$. As $\lambda$ increases the rate of convergence of $E_{PA}^{[N,N]}(\lambda)$ to the exact result becomes slower. Therefore, another faster convergent method yielding better results in the strong coupling regime needs to be considered.

## IV. Borel Resummation of the RS expansion for $E_0(\lambda)$

Borel resummation method is generally recognized as one of the most powerful methods for the summation of factorially divergent series which occur frequently in perturbation theory. The Borel transform of $E_0(\lambda)$ is defined by the contour integral [14]:

$$B(\lambda) = \frac{1}{2\pi i} \oint_C E_0(z^{-1}) z^{-1} e^{\lambda z} dz \qquad (14)$$

where $C$ is a simple closed contour encircling the origin $z = 0$. By substituting for $E_0(1/z)$ in (14) the RS perturbative expansion (9), we obtain the following series for the Borel transform:

$$B(\lambda) = \sum_{n=0}^{\infty} \frac{E_0^{(n)}}{n!} \lambda^n . \qquad (15)$$

From (10), we can show with the help of D'Alembert ratio test that the Borel series (15) has a nonzero radius of convergence $R = 2/3$. The large-order asymptotic behavior (10) and the alternating sign pattern of $E_0^{(n)}$ indicate as suggested in [7] that the closest singularity of $B(\lambda)$ to the origin is a branch point located at $\lambda = -2/3$. If $B(\lambda)$ can be analytically continued from its circle of convergence to the whole positive real axis $\text{Re}\,\lambda > 0$, then we can reconstruct $E_0(\lambda)$ from $B(\lambda)$ via the Laplace-Borel integral [7],[10],[14]:

$$E_0(\lambda) = \frac{1}{\lambda} \int_0^{\infty} B(t) e^{-t/\lambda} dt , \qquad (16)$$

provided that the integral exists for $\lambda \neq 0$. The integration variable $t$ is referred to as the Borel variable. The convergence of this integral provides a finite sum for $E_0(\lambda)$, known as the Borel sum of $E_0(\lambda)$. The evaluation of the Laplace-Borel integral requires the analytic continuation of $B(\lambda)$ beyond its circle of convergence. Therefore, we apply the diagonal Padé approximation to the power series of the Borel transform (15) to extend its domain of convergence to the whole positive real axis $\text{Re}\,\lambda > 0$. In this way, we can approximate $E_0(\lambda)$ by the Laplace-Borel integral:

$$E_B^{[N,N]}(\lambda) = \frac{1}{\lambda} \int_0^{\infty} B^{[N,N]}(t) e^{-t/\lambda} dt , \qquad (17)$$

where $B^{[N,N]}(\lambda)$ is the diagonal Padé approximant formed from the first $2N+1$ terms of the Borel series (15). Having investigated the analyticity structure of the first 25 Padé approximants $B^{[N,N]}(\lambda)$ we found that they have no poles on the positive real axis $\text{Re}\,\lambda > 0$ except for the ones with $N = 3, 7, 18, 19, 21$ and $22$. By avoiding such singular Padé approximants, we can analytically continue $B^{[N,N]}(\lambda)$ to the whole positive real axis $\text{Re}\,\lambda > 0$ and safely evaluate the Borel sum $E_B^{[N,N]}(\lambda)$, which converges to $E_0(\lambda)$ with increasing $N$.

## V. Borel-Padé Summability with Conformal Mapping

In this section, we shall combine the Borel-Padé summability technique with conformal mapping of the Borel plane [10], [14] in the resummation of the RS expansion (9) to achieve a faster convergence and to improve the accuracy. The most prominent issue in the theory of the Borel resummation is the construction of the analytic continuation for the Borel transform (15) beyond its circle of convergence. This analytic continuation can be accomplished by the conformal mapping used in [10]:

$$z(\lambda) = \frac{\sqrt{1+s\lambda} - 1}{\sqrt{1+s\lambda} + 1} , \quad \text{with} \quad s = \frac{3}{2} , \qquad (18)$$

which transforms the complex $\lambda$-plane with a cut along $\lambda \leq -1/s$, (i.e. Borel plane), onto the interior of a unit circle around the origin in the complex $z$-plane. This mapping preserves the origin and transforms $\lambda = \infty$ to $z = 1$ in such a way that places the cut in the Borel plane along the boundary of the unit circle $|z| = 1$ in the $z$-plane. For $\lambda \geq 0$, we have $0 \leq z(\lambda) < 1$. The inverse of (18) is given by:

$$\lambda(z) = \frac{4}{s} \frac{z}{(1-z)^2} . \qquad (19)$$

Now, the idea is to express Borel series (15) in terms of the conformal variable $z$ to obtain a convergent expansion [14] in $z(\lambda)$ for all $\lambda \geq 0$. By expanding $\lambda^n$ in positive powers of the conformal variable $z$ and then substituting the resulting expansion into the Borel series (15) we obtain a convergent Taylor expansion for $|z| < 1$:

$$W(z) = B(\lambda(z)) = 1 + \sum_{k=1}^{\infty} c_k z^k , \qquad (20)$$

where the coefficients $c_k$ are given by:

$$c_k = \sum_{n=1}^{k} \left(\frac{4}{s}\right)^n \frac{E_0^{(n)} (k+n-1)!}{n!\,(2n-1)!\,(k-n)!} . \qquad (21)$$

The analytic continuation of the Borel transform (15) to the whole positive real axis, $\text{Re}\,\lambda > 0$, can now be accomplished by replacing $z$ in (20) with the conformal mapping (18); obtaining:

$$W(z(\lambda)) = B(\lambda) = \sum_{k=0}^{\infty} c_k \left( \frac{\sqrt{1+s\lambda} - 1}{\sqrt{1+s\lambda} + 1} \right)^k , \quad (22)$$

with $c_0 = 1$. Given that all the singularities of the Borel transform being mapped onto the boundary of the unit circle $|z|=1$, the re-expanded series (22) provides the analytic continuation of the Borel transform (15) outside the disk of convergence $|\lambda| < 1/s$ in the Borel plane. In practice, we know only a limited number of the coefficients $c_k$. Therefore, we use the partial sum of (22) to evaluate the Laplace-Borel integral (16). But the use of this partial sum, however, spoils the accuracy of the Borel sum of $E_0(\lambda)$ for relatively large $\lambda$. To improve the accuracy and the rate of convergence of the Borel sum of $E_0(\lambda)$ we consider instead resumming the partial sum of (20) by a diagonal Padé approximant, which we denote by $W^{[N,N]}(z)$. Then, we replace the Borel transform in (16) with the Padé approximant $W^{[N,N]}(z)$; where $z$ is given by (18). In this way, we obtain the Borel sum of $E_0(\lambda)$ as:

$$E_{CM}^{[N,N]}(\lambda) = \frac{4}{\lambda s} \int_0^1 W^{[N,N]}(t) F(\lambda, t) \, dt , \quad (23)$$

where $F(\lambda, t)$ is given by:

$$F(\lambda, t) = \frac{1+t}{(1-t)^3} e^{-4t/\lambda s(1-t)^2} . \quad (24)$$

By investigating the analyticity structure of the Padé approximants $W^{[N,N]}(t)$ up to $N = 25$ we found that for $N = 2, 15, 17, 20$ and $23$ the relevant Padé approximants introduce unphysical poles in the interval $0 < t < 1$, making the integral in (23) impossible to evaluate without considering the Cauchy principal value. Therefore, we will avoid using these Padé approximants in our calculations. With increasing $N$, our numerical computations show that the Borel sum $E_{CM}^{[N,N]}(\lambda)$ converges to $E_0(\lambda)$ faster than $E_B^{[N,N]}(\lambda)$ especially for large $\lambda$.

## VI. NUMERICAL RESULTS

In this section, we present the numerical results based on the resummation methods introduced earlier. Using the first 50 coefficients in the RS perturbative expansion for $E_0(\lambda)$, we have computed with Maple the first 25 diagonal Padé approximants for the RS expansion (9), Borel series (15) and the Borel transform expansion with conformal mapping (20), obtaining $E_{PA}^{[N,N]}(\lambda)$, $B^{[N,N]}(\lambda)$ and $W^{[N,N]}(z)$. Then, we calculated the zeros of the denominator polynomials of the Padé approximants $B^{[N,N]}(\lambda)$ and $W^{[N,N]}(z)$ for $N = 1, 2, \ldots, 25$ and determined the simple poles of these approximants. By investigating the locations of these poles, we found that the poles for $B^{[N,N]}(\lambda)$ all lie outside the integration domain of the Laplace-Borel integral (17) except for $N = 3, 7, 18, 19, 21$ and $22$, and the poles for $W^{[N,N]}(t)$ all lie outside the integration domain $0 \leq t < 1$ in (23) except for $N = 2, 15, 17, 20$ and $23$. In this way, we identified the Padé approximants with no poles on the relevant integration domains as the proper class of approximants to be used in our calculations. Using our proper Padé approximants in (17) and (23), we managed to evaluate numerically the relevant Borel sums $E_B^{[N,N]}(\lambda)$ and $E_{CM}^{[N,N]}(\lambda)$ of the divergent RS expansion for ground state energy of the quartic anharmonic oscillator.

In our calculations, we have used Maple to do the relevant numerical computations with a high decimal precision (up to 300 decimal places). In Tables I and II we illustrate the rate of convergence of $E_{PA}^{[N,N]}(\lambda)$, $E_B^{[N,N]}(\lambda)$ and $E_{CM}^{[N,N]}(\lambda)$ to the exact results (up to 15 significant figures) [17]-[21] as well as the rate of divergence of $E_{RS}^{[N,N]}(\lambda)$ with increasing $N$ for $\lambda = 0.2, 1, 4$ and $100$, where $2N$ is the order of perturbation. As seen from the tables below, the Borel sums with conformal mapping $E_{CM}^{[N,N]}(\lambda)$ converge relatively faster to the exact results with increasing $N$ and this becomes more apparent for larger values of $\lambda$. Table II shows that the rate of convergence of Padé approximation for relatively large values of $\lambda$ is remarkably slower than Borel's Method of Summability. In Table III, we compare the results obtained from Padé approximation, Borel resummation and the Borel resummation with conformal mapping with the exact results available to 10 decimal places for $\lambda$ between 0.1 and 2000. For coupling constants $\lambda \leq 0.4$, Table III shows that $E_{PA}^{[N,N]}(\lambda)$, $E_B^{[N,N]}(\lambda)$ and $E_{CM}^{[N,N]}(\lambda)$ yield the exact results to 10 decimal places whereas for $\lambda \geq 4$ we can see from the table that Padé approximation and Borel resummation do not sum RS perturbative expansion as effectively as the Borel resummation with conformal mapping (23). For coupling constants as large as $\lambda = 2000$, Table III shows in percentage terms that the Borel resummation with conformal mapping reproduces the exact ground state energy with an error less than one percent whereas the Padé approximation and the standard Borel resummation produce errors in the results greater than 70 percent and 26 percent respectively.

**Table I:** Rate of convergence of Padé approximation and Borel resummation, with and without conformal mapping, with increasing order of perturbation and the rate of divergence of RS perturbative expansion for $\lambda = 0.2$ and $\lambda = 1$.

| $\lambda = 0.2$ | | | | |
|---|---|---|---|---|
| $N$ | $E_{RS}^{[2N]}(\lambda)$ | $E_{PA}^{[N,N]}(\lambda)$ | $E_{B}^{[N,N]}(\lambda)$ | $E_{CM}^{[N,N]}(\lambda)$ |
| 1 | 1.09750000000000 | 1.11111111111111 | 1.11434921525114 | 1.11576568994656 |
| 4 | 0.67150362147644 | 1.11827272295579 | 1.11829223203545 | 1.11829271491758 |
| 5 | -2.55720131255401 | 1.11828840520697 | 1.11829261800686 | 1.11829264236124 |
| 6 | -42.2480889195723 | 1.11829163112845 | 1.11829265169967 | 1.11829265447894 |
| 8 | -14719.0500060434 | 1.11829257635767 | 1.11829265432581 | 1.11829265436737 |
| 9 | $-3.951608936 \times 10^5$ | 1.11829263040450 | 1.11829265435845 | 1.11829265436637 |
| 10 | $-1.316486008 \times 10^7$ | 1.11829264657358 | 1.11829265436719 | 1.11829265436704 |
| 11 | $-5.332986348 \times 10^8$ | 1.11829265170351 | 1.11829265436694 | 1.11829265436703 |
| 12 | $-2.582537687 \times 10^{10}$ | 1.11829265341622 | 1.11829265436700 | 1.11829265436704 |
| 13 | $-1.473688040 \times 10^{12}$ | 1.11829265401424 | 1.11829265436704 | 1.11829265436704 |
| 14 | $-9.788558688 \times 10^{13}$ | 1.11829265423153 | 1.11829265436704 | 1.11829265436704 |
| 16 | $-6.537405037 \times 10^{17}$ | 1.11829265434515 | 1.11829265436704 | 1.11829265436704 |
| 24 | $-1.138802608 \times 10^{35}$ | 1.11829265436700 | 1.11829265436704 | 1.11829265436704 |
| 25 | $-2.470062516 \times 10^{37}$ | 1.11829265436702 | 1.11829265436704 | 1.11829265436704 |
| Exact | 1.11829265436704 | 1.11829265436704 | 1.11829265436704 | 1.11829265436704 |
| $\lambda = 1$ | | | | |
| $N$ | $E_{RS}^{[2N]}(\lambda)$ | $E_{PA}^{[N,N]}(\lambda)$ | $E_{B}^{[N,N]}(\lambda)$ | $E_{CM}^{[N,N]}(\lambda)$ |
| 1 | 0.4375000000000 | 1.27272727272727 | 1.32409107378417 | 1.34665644313731 |
| 4 | $-2.267122600 \times 10^5$ | 1.38375649722835 | 1.39191813054962 | 1.39237485133172 |
| 5 | $-4.478112143 \times 10^7$ | 1.38807560338964 | 1.39224589319169 | 1.39232943567057 |
| 6 | $-1.281827249 \times 10^{10}$ | 1.39010375465198 | 1.39232661366795 | 1.39235142023652 |
| 8 | $-2.608777784 \times 10^{15}$ | 1.39164801814861 | 1.39234917447971 | 1.39235160832607 |
| 9 | $-1.725031026 \times 10^{18}$ | 1.39193836533906 | 1.39235060809535 | 1.39235157173283 |
| 10 | $-1.419298922 \times 10^{21}$ | 1.39210249509863 | 1.39235162941168 | 1.39235164042288 |
| 11 | $-1.422760306 \times 10^{24}$ | 1.39219801005589 | 1.39235155723875 | 1.39235163750826 |
| 12 | $-1.707599422 \times 10^{27}$ | 1.39225501035586 | 1.39235159132195 | 1.39235163980814 |
| 13 | $-2.417999934 \times 10^{30}$ | 1.39228978425356 | 1.39235164497134 | 1.39235164157819 |
| 14 | $-3.989416229 \times 10^{33}$ | 1.39231141653459 | 1.39235163679496 | 1.39235164143754 |
| 16 | $-1.647639017 \times 10^{40}$ | 1.39233391548343 | 1.39235164148149 | 1.39235164152224 |
| 24 | $-4.269314309 \times 10^{68}$ | 1.39235063782670 | 1.39235164152952 | 1.39235164153029 |
| 25 | $-2.310225366 \times 10^{72}$ | 1.39235091567726 | 1.39235164153000 | 1.39235164153029 |
| Exact | 1.39235164153029 | 1.39235164153029 | 1.39235164153029 | 1.39235164153029 |

**Table II**: Rate of convergence of Padé approximation and Borel resummation, with and without conformal mapping, with increasing order of perturbation and the rate of divergence of RS perturbative expansion for $\lambda = 4$ and $\lambda = 100$.

| $\lambda = 4$ | | | | |
|---|---|---|---|---|
| $N$ | $E_{RS}^{[2N]}(\lambda)$ | $E_{PA}^{[N,N]}(\lambda)$ | $E_{B}^{[N,N]}(\lambda)$ | $E_{CM}^{[N,N]}(\lambda)$ |
| 1 | -17.000000000000 | 1.37500000000000 | 1.55012871078021 | 1.64801395937678 |
| 4 | $-1.576904211 \times 10^{10}$ | 1.74458656716086 | 1.88354836978232 | 1.90332943529492 |
| 5 | $-4.929509300 \times 10^{13}$ | 1.78921784846164 | 1.89475286122881 | 1.90171934465575 |
| 6 | $-2.240522467 \times 10^{17}$ | 1.81940831749416 | 1.89955030367304 | 1.90302250543321 |
| 8 | $-1.155809879 \times 10^{25}$ | 1.85543521579422 | 1.90223791004483 | 1.90311331193473 |
| 9 | $-1.218692779 \times 10^{29}$ | 1.86635179542483 | 1.90260662212985 | 1.90310376806957 |
| 10 | $-1.599943152 \times 10^{33}$ | 1.87443396856257 | 1.90307401843857 | 1.90313398024850 |
| 11 | $-2.560393982 \times 10^{37}$ | 1.88050705986074 | 1.90302124533212 | 1.90313168329998 |
| 12 | $-4.907557732 \times 10^{41}$ | 1.88513017858614 | 1.90305121465019 | 1.90313394475166 |
| 13 | $-1.110104865 \times 10^{46}$ | 1.88869021609447 | 1.90313726688144 | 1.90313699721225 |
| 14 | $-2.926464346 \times 10^{50}$ | 1.89145990523098 | 1.90311676797482 | 1.90313655278772 |
| 16 | $-3.087233090 \times 10^{59}$ | 1.89535663206059 | 1.90313502117437 | 1.90313685582456 |
| 24 | $-3.417944132 \times 10^{97}$ | 1.90127036112281 | 1.90313684731966 | 1.90313694464263 |
| 25 | $-2.958013301 \times 10^{102}$ | 1.90154860683346 | 1.90313689209695 | 1.90313694541205 |
| Exact | 1.90313694545900 | 1.90313694545900 | 1.90313694545900 | 1.90313694545900 |
| $\lambda = 100$ | | | | |
| $N$ | $E_{RS}^{[2N]}(\lambda)$ | $E_{PA}^{[N,N]}(\lambda)$ | $E_{B}^{[N,N]}(\lambda)$ | $E_{CM}^{[N,N]}(\lambda)$ |
| 1 | -13049.00000000 | 1.42613636363636 | 1.81844199401582 | 2.23071877627509 |
| 4 | $-2.454816061 \times 10^{21}$ | 2.10680544605624 | 3.41991191234569 | 4.82236712582861 |
| 5 | $-4.777789539 \times 10^{27}$ | 2.25897130728198 | 3.72910546989454 | 4.67102031713377 |
| 6 | $-1.353662113 \times 10^{34}$ | 2.39186398141383 | 3.97802447819192 | 4.87729411100690 |
| 8 | $-2.718597976 \times 10^{47}$ | 2.61636827732774 | 4.29221483915724 | 4.93734838139868 |
| 9 | $-1.789532587 \times 10^{54}$ | 2.71314145350933 | 4.38786906319967 | 4.92981352210877 |
| 10 | $-1.467014788 \times 10^{61}$ | 2.80188407550551 | 4.65083753220694 | 4.97270949792091 |
| 11 | $-1.466199403 \times 10^{68}$ | 2.88379532882792 | 4.59978514668735 | 4.96643585645683 |
| 12 | $-1.755343011 \times 10^{75}$ | 2.95981189824916 | 4.63147763949178 | 4.97371692630670 |
| 13 | $-2.480347698 \times 10^{82}$ | 3.03068138767302 | 4.84084086007409 | 4.99708496397132 |
| 14 | $-4.084853819 \times 10^{89}$ | 3.09701140608162 | 4.75724107310096 | 4.98952978854829 |
| 16 | $-1.682076714 \times 10^{104}$ | 3.21797707307754 | 4.87436956034579 | 4.99342654192674 |
| 24 | $-4.328571015 \times 10^{164}$ | 3.59094029559601 | 4.95043489249390 | 4.99865751756073 |
| 25 | $-2.340997539 \times 10^{172}$ | 3.62843617084790 | 4.95883674037978 | 4.99924243571278 |
| Exact | 4.99941754513759 | 4.99941754513759 | 4.99941754513759 | 4.99941754513759 |

Table III: Comparison of Padé approximation and Borel resummation, with and without conformal mapping, with the exact energy results [17]-[21] for the quartic anharmonic oscillator at different values of the coupling constant.

| $\lambda$ | $E_{PA}^{[25,25]}(\lambda)$ | Error (%) | $E_{B}^{[25,25]}(\lambda)$ | Error (%) | $E_{CM}^{[25,25]}(\lambda)$ | Error (%) | Exact |
|---|---|---|---|---|---|---|---|
| 0.10 | 1.065285510 | 0.00000000 | 1.065285510 | 0.00000000 | 1.065285510 | 0.00000000 | 1.065285510 |
| 0.15 | 1.092905010 | 0.00000000 | 1.092905010 | 0.00000000 | 1.092905010 | 0.00000000 | 1.092905010 |
| 0.20 | 1.118292654 | 0.00000000 | 1.118292654 | 0.00000000 | 1.118292654 | 0.00000000 | 1.118292654 |
| 0.25 | 1.141901840 | 0.00000000 | 1.141901840 | 0.00000000 | 1.141901840 | 0.00000000 | 1.141901840 |
| 0.30 | 1.164047157 | 0.00000000 | 1.164047157 | 0.00000000 | 1.164047157 | 0.00000000 | 1.164047157 |
| 0.35 | 1.184958484 | 0.00000000 | 1.184958484 | 0.00000000 | 1.184958484 | 0.00000000 | 1.184958484 |
| 0.40 | 1.204810327 | 0.00000000 | 1.204810327 | 0.00000000 | 1.204810327 | 0.00000000 | 1.204810327 |
| 0.45 | 1.223739118 | $8.172 \times 10^{-8}$ | 1.223739119 | 0.00000000 | 1.223739119 | 0.00000000 | 1.223739119 |
| 0.50 | 1.241854058 | $1.610 \times 10^{-7}$ | 1.241854060 | 0.00000000 | 1.241854060 | 0.00000000 | 1.241854060 |
| 1 | 1.392350916 | 0.00005214 | 1.392351642 | 0.00000000 | 1.392351642 | 0.00000000 | 1.392351642 |
| 2 | 1.607479513 | 0.00384370 | 1.607541302 | 0.00000000 | 1.607541302 | 0.00000000 | 1.607541302 |
| 4 | 1.901548607 | 0.08345894 | 1.903136892 | 0.00000278 | 1.903136945 | 0.00000000 | 1.903136945 |
| 10 | 2.417565253 | 1.29059095 | 2.449162086 | 0.00048939 | 2.449174057 | $6.125 \times 10^{-7}$ | 2.449174072 |
| 100 | 3.628436171 | 27.4228220 | 4.958836740 | 0.81171066 | 4.999242436 | 0.00350259 | 4.999417545 |
| 400 | 3.877051655 | 50.6853298 | 7.313893098 | 6.96997140 | 7.855794320 | 0.07718728 | 7.861862678 |
| 2000 | 3.953407375 | 70.4714898 | 9.861171928 | 26.3456334 | 13.265159677 | 0.92080936 | 13.388441701 |

## VII. CONCLUSION

We have shown that a combination of an analytic continuation of the Borel plane via conformal mapping and a subsequent resummation of the Borel transform by Padé approximants lead to a significant acceleration of the convergence of the resummed RS perturbative series for the ground state energy of the quartic anharmonic oscillator. The technique of the conformal mapping of the Borel plane refines the Borel resummation and improves its accuracy. This technique gives good predictions in the strong coupling domain where perturbation theory is not applicable. Our results indicates that the divergent RS perturbative expansion for $E_0(\lambda)$ is Borel summable to the exact results for coupling constants as large as $\lambda = 2000$ under the conformal mapping of the Borel plane. Also, we conclude that the Borel-Padé summability when combined with the conformal mapping of the Borel plane can produce reasonably accurate results from RS-type perturbative expansions of physical observables even in moderately strong coupling regimes. We add also that there is no problem in applying the same method we used here to excited states of the quartic anharmonic oscillator.

## APPENDIX

Below we list the coefficients $E_0^{(n)}$ of the first 30 correction terms in the Rayleigh-Schrodinger perturbation series for the ground state energy of the quartic anharmonic oscillator.

Table IV: The coefficients $E_0^{(n)}$ of the first 30 correction terms in the Rayleigh-Schrodinger perturbation series for the ground state energy of the quartic anharmonic oscillator.

| $n$ | $E_0^{(n)}$ | $n$ | $E_0^{(n)}$ |
|---|---|---|---|
| 1 | $\frac{3}{4}$ | 16 | $-\frac{19138592785256092788782808 4605}{70368744177664}$ |
| 2 | $-\frac{21}{16}$ | 17 | $\frac{190806107833206980489642266 01511}{281474976710656}$ |
| 3 | $\frac{333}{64}$ | 18 | $-\frac{4031194983593309788607032686 292335}{2251799813685248}$ |
| 4 | $-\frac{30885}{1024}$ | 19 | $\frac{44982060454076583616052969749 1458635}{9007199254740992}$ |
| 5 | $\frac{916731}{4096}$ | 20 | $-\frac{2114910575845607954251483096 63914344715}{144115188075855872}$ |
| 6 | $-\frac{65518401}{32768}$ | 21 | $\frac{26120222383762781149654970754 9344170 34805}{576460752303423488}$ |
| 7 | $\frac{2723294673}{131072}$ | 22 | $-\frac{676374022879448974059639416182 2714805728135}{4611686018427387904}$ |
| 8 | $-\frac{1030495099053}{4194304}$ | 23 | $\frac{9161175277997465779846418004843 28417887454935}{18446744073709551616}$ |
| 9 | $\frac{54626982511455}{16777216}$ | 24 | $-\frac{10364646961057201298031882166199 0185513 1610583785}{59029581035870565 1712}$ |
| 10 | $-\frac{6417007431590595}{134217728}$ | 25 | $\frac{15277377494984443898348201883095 70223243 92611587683}{2361183241434822606848}$ |
| 11 | $\frac{413837985580636167}{536870912}$ | 26 | $-\frac{46864661934282542604967271035084 21202836 4938136115903}{18889465931478580854784}$ |
| 12 | $-\frac{116344863173284543665}{8589934592}$ | 27 | $\frac{74684247714146341178454091235341 19931849 505977965315619}{75557863725914323419136}$ |
| 13 | $\frac{8855406003085477228503}{34359738368}$ | 28 | $-\frac{49394798987776425687768816380486 10516053 5783672500 19222857}{120892581961462917470 6176}$ |
| 14 | $-\frac{1451836748576538293163705}{274877906944}$ | 29 | $\frac{84628477407456510683723784862205 32150621 53108475654677813615}{483570327845851669882 4704}$ |
| 15 | $\frac{127561682802713500067360049}{1099511627776}$ | 30 | $-\frac{30012182458330101215997089701342 88492619 73957431788548963467657}{38685626227668133590597632}$ |